\documentclass[twocolumn]{jpsj2}
\usepackage{dcolumn}% Align table columns on decimal point
\usepackage{bm}% bold math

\title{Polaronic States with Spin-Charge-Coupled Excitation in a One-Dimensional Dimerized Mott Insulator K-TCNQ}
\author{
Nobuya \textsc{Maeshima}${}^{1,2}$\thanks{ Present address: Institute of Materials Science, University of Tsukuba, Tsukuba 305-8573} and
Kenji \textsc{Yonemitsu}${}^{2,3}$
}
\inst{
${}^1$ Department of Chemistry, Tohoku University, Aramaki, Aoba-ku, Sendai 980-8578\\
${}^2$ Institute for Molecular Science, Okazaki 444-8585 \\
${}^3$ Department of Functional Molecular Science, Graduate University for Advanced Studies, Okazaki 444-8585 
}

\recdate{\today}

\abst{
We discuss photogenerated midgap states of a one-dimensional (1D) dimerized Mott insulator, potassium-tetracyanoquinodimethane (K-TCNQ).  Two types of phonon modes are taken into account: intermolecular and intramolecular vibrations.  We treat these phonon modes adiabatically and analyze a theoretical model by using the density-matrix renormalization group (DMRG).  Our numerical results demonstrate that the intermolecular lattice distortion is necessary to reproduce the photoinduced midgap absorption in K-TCNQ.  We find two types of midgap states.  One is a usual polaronic state characterized by a localized elementary excitation.  The other is superposition of two types of excitations, a doped-carrier state and a triplet-dimer state, which can be generally observed in 1D dimerized Mott insulators, not limited to K-TCNQ.
}

\kword{strongly correlated electron systems, electron-phonon coupling, photoinduced phase transition, polaron, density matrix renormalization group}
\begin{document}
\sloppy
\maketitle

\section{Introduction}

%<Phenomena>
Midgap states observed in one-dimensional (1D) organic band insulators are due to excited states that consist of localized carriers coupled with phonons.~\cite{heeger,okamoto1}  A polaron state is a typical example of these midgap states and has attracted a lot of research interest.  In particular, photogeneration of the polaron state has been intensively studied for several decades in the context of ultrafast generation of a polaron and its relaxation dynamics.

%<Why is it important>
To understand the nature of these midgap states also provides us a good insight into the fundamental mechanism of macroscopic photoinduced phenomena, called photoinduced phase transitions.~\cite{nasu,yonemitsu1}  Currently, considerable attention has been paid to the photoinduced phase transitions, in particular, of strongly correlated electron systems covering inorganic~\cite{yu,PrMnO3,VO2,TaS2} and organic~\cite{TTFCA1,TTFCA2,poly,MX,MMX,EDO,sitaET,KTCNQ1,KTCNQ2,KTCNQ3,TTTA1,takeda} materials including those which show photoinduced inverse spin-Peierls transitions.~\cite{KTCNQ1,KTCNQ2,KTCNQ3,TTTA1,takeda}  Although the transitions intrinsically accompany large-scale changes, their initial seeds would be local and microscopic ones~\cite{nasu,yonemitsu1} that are likely to be observed through the appearance of midgap states immediately after photoirradiation.  Therefore, to examine the midgap states can help us to understand the starting point of the photoinduced phase transitions.~\cite{matsueda}

%<Which phenomena do you focus on?>
In this study we focus on an organic material, potassium-tetracyanoquinodimethane (K-TCNQ).  This material is classified as a 1D dimerized Mott insulator below the spin-Peierls transition temperature, $T_{\rm sP}=395$K.~\cite{vegter,sakai,terauchi}  In 1991, Koshihara {\it et al.} have demonstrated that an irradiation of a pulsed laser weakens the lattice dimerization of K-TCNQ in the dimerized phase, which is regarded as a photoinduced inverse spin-Peierls transition.~\cite{KTCNQ1}  More recently, a high-resolution experiment has revealed several ultra-fast phenomena immediately after the photoirradiation, for example, the emergence of a midgap peak around 0.3 eV in the reflectivity spectrum within the time resolution of 150fs.~\cite{KTCNQ2,KTCNQ3} The decrease of the lattice dimerization is found to occur after the midgap state appears.

%<our purpose>
The purpose of our study is to elucidate the nature of the midgap state in K-TCNQ.  Our preceding study has made a suggestion that the midgap state is due to a purely electronic excitation of photoinduced {\it mobile} carriers.~\cite{maeshima3}  This scenario is excellent in that the instant appearance of the midgap peak is explained naturally.  However it is inconsistent with the experimental fact that the Drude component caused by the carriers does not appear.~\cite{KTCNQ2}

%<point>
In this paper, we discuss an alternative mechanism; polaronic localized midgap states generated by the photoexcitation.  We take account of two types of phonons, intermolecular and intramolecular vibrations, and treat these phonon modes adiabatically to analyze a theoretical model by using the density-matrix renormalization group (DMRG).~\cite{white,hallberg,kuhner,jeckelmann}
Our numerical results demonstrate that the midgap state in K-TCNQ can be reproduced by two polaronic lattice configurations.  One is a configuration with the relaxed inter-molecular vibration mode and the unrelaxed intramolecular vibration mode.  This configuration provides a shoulder structure around 0.1eV in addition to the main midgap peak at 0.3 eV.
The other configuration is accompanied with the relaxed intermolecular and intramolecular vibration modes,  where the shoulder and the peak merge into a single peak structure at 0.3 eV.
We clarify the origins of these peak/shoulder structures and find out that spin excitations play an important role, which is characteristic to 1D dimerized Mott insulators.
%We have confirmed that that the shoulder is originated from the convensional polaronic senario; the optical transition from the lowest bound state to a higher excited state.  By contrast, the main peak is shown to corresponds to 
%This paper is organized as follows.
%By contrast, the remaing lattice configuration that has relaxed intramolecular vibration and unrelaxed intermolecular vibration (relaxed $v$ and fixed $u$) is less plausible because additional peaks around 0.6eV must appear.

% coupling, the Peierls type and the Holstein type.
% to take account of lattice dimerization and intramolecular oscilation.
% large-size simulation using DMRG
% usual DMRG     to determine the stable lattice configuration
% dynamical DMRG to obtain optical absorption spectra.
% which can influence the subsequent dynamics of K-TCNQ
%<result>
%--------------------------------------------------------------------

\section{Model and Method}

\subsection{Model}

%<introduce model>
In this work, we use the 1D extended Hubbard model with Peierls and Holstein types of electron-phonon (e-ph) couplings.  The Hamiltonian is given by
 \begin{eqnarray}
{\cal H} &=& -\sum_{l=0}^{N-2}(t_0-\alpha u_l) \hat{t}_{l,l+1} +  U\sum_{l=0}^{N-1} n_{l,\uparrow}n_{l,\downarrow}\nonumber \\
&+& V\sum_{l=0}^{N-2} n_{l}n_{l+1}  -\beta\sum_{l=0}^{N-1} v_l (n_l-1) \nonumber \\
&+& \frac{K_\alpha}{2}\sum_{l=0}^{N-2} u_l^2 + \frac{K_\beta}{2}\sum_{l=0}^{N-1} v_l^2 +\Gamma\sum_{l=0}^{N-2} u_l,
\label{eq_ham}
\end{eqnarray}
where
\begin{equation}
\hat{t}_{l,l+1} = \sum_{\sigma} ( c^\dagger_{l+1,\sigma}c_{l,\sigma} +  c^\dagger_{l,\sigma}c_{l+1,\sigma} ),
\end{equation}
and $c^{\dagger}_{l,\sigma}$ ($c_{l,\sigma}$) creates (annihilates) an electron with spin $\sigma$ in the LUMO of a TCNQ molecule on site $l$, and $n_{l,\sigma}=c^{\dagger}_{l,\sigma}c_{l,\sigma}$.  The Coulomb repulsion is taken into account up to the nearest neighbor: the on-site Coulomb interaction $U$ and the nearest-neighbor interaction $V$.  We impose the open boundary condition (OBC) and $N$ denotes the system size.  The parameter $t_0$ gives the transfer integral of the uniform lattice.
The distortion of the $l$-th bond is denoted by $u_l$, and the molecular deformation is given by $v_l$.  The e-ph coupling constants are given by $\alpha$ and $\beta$, and the elastic spring constants are denoted by $K_\alpha$ and $K_\beta$, respectively.  Taking the adiabatic approximation for phonons, we omit their kinetic energy terms.  $\Gamma$ is introduced to keep the chain length the same as that of the unrelaxed configuration, i.e.,
\begin{equation}
\sum_{l=0}^{N-2} u_l=0.
\label{eq_const}
\end{equation}
For simplicity, we re-define phonon variables in the following; $\alpha u_l \to u_l$ and $\beta v_l \to v_l$.  Accordingly, the elastic constants are renormalized as $K_\alpha/\alpha^2 \to K_\alpha$ and $K_\beta/\beta^2 \to K_\beta$.

\subsection{Method}
To obtain the stable (i.e., fully relaxed) lattice configuration, we use the Hellmann-Feynman theorem;
\begin{equation}
 \frac{\partial \langle {\cal H}\rangle}{\partial u_l} = 0 \quad {\rm and/or} \quad
 \frac{\partial \langle {\cal H}\rangle}{\partial v_l} = 0 \quad {\rm for \ all} \quad l.
\end{equation}
These equations lead to the following relations,
\begin{eqnarray}
 u_l &=& (\langle \hat{t}_{l,l+1}\rangle+\Gamma)/K_\alpha, \quad {\rm and } \label{eq_lat1} \\
 v_l &=& (\langle n_l \rangle -1)/K_\beta,
\label{eq_lat2}
\end{eqnarray}
where $\langle \rangle$ is the expectation value for the eigenstate we obtained, and $\Gamma$ is derived as
\begin{equation}
 \Gamma= -\sum_{l=0}^{N-2}\langle \hat{t}_{l,l+1}\rangle/(N-1)
\label{eq_gamma}
\end{equation}
from eq.~(\ref{eq_const}).  We evaluate the RHSs of eqs.~(\ref{eq_lat1}),~(\ref{eq_lat2}), and ~(\ref{eq_gamma}) by using the DMRG method for a certain lattice configuration $(\bm{u},\bm{v})$ and $\Gamma$, and then we obtain new values of $(\bm{u},\bm{v})$ and $\Gamma$.
The new configuration and $\Gamma$ are used to evaluate the RHSs of eqs.~(\ref{eq_lat1}),~(\ref{eq_lat2}), and~(\ref{eq_gamma}) again and we obtain next values.  We execute this iterative procedure and finally reach the stable lattice configuration.  For the final configuration, we apply the dynamical DMRG (DDMRG)~\cite{hallberg,kuhner,jeckelmann} to calculate the spectral function given by~\cite{shastry}
\begin{equation}
\chi(\omega)= \frac{\pi}{N} \sum_i |\langle\psi_i|J|\psi_0\rangle|^2 \delta(\omega+E_{0}-E_i),
\label{eq_sigma-reg}
\end{equation}
where
\begin{equation}
J\equiv i\sum_{l,\sigma}(t_0-u_l)( c^\dagger_{l+1,\sigma}c_{l,\sigma} -  c^\dagger_{l,\sigma}c_{l+1,\sigma})
\end{equation}
is the current operator, $|\psi_i\rangle$ the $i$-th eigenstate, and $E_i$ the corresponding energy.  $\chi(\omega)$ is related to the optical conductivity spectrum as $\chi(\omega)=\omega\sigma(\omega)$.  The lattice constant is set to be unity for simplicity. % We note that the Drude part is omitted because the model~(\ref{eq_ham}) is treated only in the insulating phase.
In the actual computation, Lorentzians with the broadening parameter $\epsilon$ are substituted for the $\delta$ functions in eq.~(\ref{eq_sigma-reg}).   We have carried out almost all DMRG calculations on $N=40$ systems with retained bases $m=80$.

%----------------------------------------------------------
\section{Optical Conductivity at Half Filling}

%<physical parameters>
Before discussing polaronic midgap states in doped systems, we carry out calculation at half filling to obtain model parameters, $U$,$V$, and $K_\alpha$ for K-TCNQ.
$K_\beta$ is not determined in this study, as noted in \S 4.
Our method of obtaining these parameters is as follows.
First, we find the most plausible values of $U$ and $V$ by comparing the calculated optical conductivity $\sigma(\omega)$ with the experimental result by Yakushi {\it et al.}~\cite{yakushi}. 
Here the bond distortion is fixed to $u_l=(-1)^l \delta t_0$, where $\delta$ is the dimerization parameter and estimated as $\delta=0.46(\equiv \delta_{\rm e})$ from the extended H\"{u}ckel calculation by K. Ikegami {\it et al.}~\cite{KTCNQ3}.
Thus, at this stage, we do not use the iterative procedure described in \S 2.2.

\begin{figure}[hbt]
\begin{center}
\includegraphics[width=6.0cm,clip]{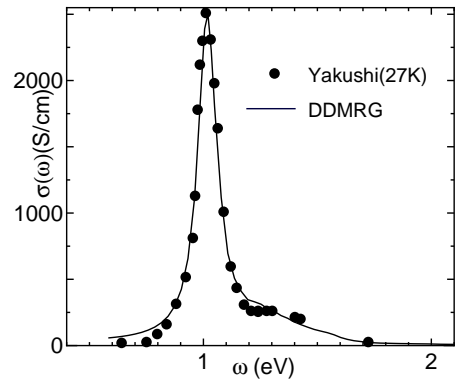}
\end{center}
\caption{Optical conductivity spectra of K-TCNQ. The symbols show experimental result at 27 K in ref.~\citen{yakushi}, and the solid line is our DDMRG result for $N=40$.}
\label{fig_yakushi}
\end{figure}

Our best DDMRG result of $\sigma(\omega)$ and the experimental data by Yakushi {\it et al.}~\cite{yakushi} are shown in Fig.~\ref{fig_yakushi}.  The DDMRG result is calculated for $U/t_0=5$, $V/t_0=1.7$, and $t_0=0.291{\rm eV}$, with the broadening parameter $\epsilon=0.16t_0$.  As for the on-site Coulomb interaction $U$, our estimation is comparable to the values in the previous studies.~\cite{yakushi,meneghetti,gallagher}   In addition, the relatively large $V$ is consistent with our previous conclusion that large $V$ is necessary to reproduce the sharp lowest peak and the broad higher-energy shoulder,~\cite{maeshima4} which are characteristic to K-TCNQ.

Then we determine $K_{\alpha}$ realizing $\delta=\delta_{\rm e}$ as the stable lattice configuration at half filling to be $K_\alpha=1.6/t_0$.  Actually, the obtained stable lattice configuration at half filling with OBC is not completely uniformly dimerized.  Besides a small boundary effect due to the OBC, there is a small difference between even and odd bonds: the number of even bonds is larger than that of odd bonds by one.  
Hence, we obtain the relaxed lattice configuration where the average of dimerization is equal to $\delta_{\rm e}$;
\begin{equation}
  \frac{1}{N-1}\sum_{l=0}^{N-2} (-1)^l u_l/t_0 = \delta_e,
\end{equation}
 and then we confirm that $\sigma(\omega)$ for this stable configuration is almost identical to that for the uniformly dimerized configuration.

\section{Polaronic States}

Now, we discuss photogenerated polaronic states.
In general, photoexcitation with energy corresponding to the charge transfer (CT) excitation generates two carriers.  For 1D Mott insulators, these are known as a holon and a doublon.~\cite{mizuno}  In the experiment by H. Okamoto {\it et al.}, the pump excitation with 1.55eV is considered to generate mobile holons and doublons.~\cite{KTCNQ2} These carriers move freely and are made separate within the time scale of $\hbar/t_0\sim 10^{-14}$ sec, which is much shorter than the experimental time resolution ($\sim 150$fs)~\cite{KTCNQ2} and the phonon dynamics.
Hence, we consider that the lattice relaxation starts after the carriers are well-separated, suggesting the interaction effect between the carriers is negligible.
Because of the particle-hole symmetry of this model, we consider the case where only one doublon exists on the system.  We confirm that the case with a single holon shows the same results.  As for the type of lattice relaxation, we consider the following cases:  ({\it case 1}) $u$-relaxed and $v$-fixed, ({\it case 2}) $u$- and $v$-relaxed, and ({\it case 3}) $u$-fixed and $v$-relaxed, where $u(v)$-relaxed means that the intermolecular (intramolecular) distortion $u(v)$ is relaxed to the stable inhomogeneous configuration, and the $u(v)$-fixed denotes that the lattice configuration is fixed to the homogeneous one at half filling.

\begin{figure}[hbt]
\begin{center}
\includegraphics[width=6.0cm,clip]{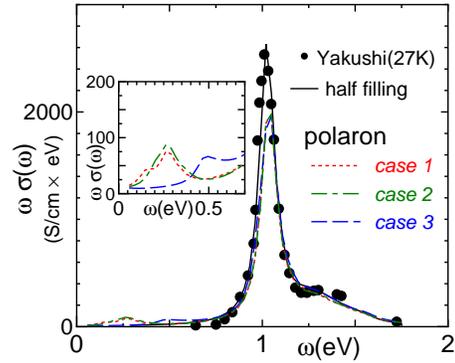}
\end{center}
\caption{(Color online) Spectral functions for polaronic states where a single electron is introduced to $N=40$ systems.  In {\it cases 2} and {\it 3}, $K_\beta$ is set to $1.0/t_0$.  The inset shows the magnified picture for the midgap region. }
\label{fig_polsw}
\end{figure}

Figure~\ref{fig_polsw} shows $\omega\sigma(\omega)$ of the polaron states for these three lattice configurations. 
The lowest-energy peak due to the open boundary condition is invisible in this and following figures.
The results suggest that both of {\it case 1} and {\it case 2} well reproduce the midgap peak at 0.3 eV, although a small difference exists between them.  The former shows a shoulder structure on the lower energy side of the main midgap peak. By contrast, {\it case 2} seems to exhibit a single peak at 0.3eV.   In {\it case 3}, there appears a midgap state above 0.5eV, being inconsistent with the experimental result.  These observations suggest that the relaxation of $u$ is necessary to reproduce the experimental result.

%\textcolor{blue}{We also note that the Drude ``precursor'',~\cite{fye} the metallic low-energy component characterisitc to open-boundary systems, is invisible.  This is because of (i) low concentration of the carriers (one electron for 40 site in our calculation), and (ii) the inhomogeneity of the system caused by the polaron formation.}

The remaining problem is whether the relaxation of $v$ is necessary or not. In fact, we cannot reach the clear answer at present.
The experimental resolution is not high enough to decide whether or not the low-energy shoulder is realized and to determine the plausible value of $K_\beta$.
% The limit of experimental resolution disables us to find out whether the lower-energy shoulder is realized or not.  \textcolor{red}{Hence, we do not  answer the question of whether the intramolecular mode is necessary or not, much less determine the value of $K_\beta$.}
In the following, we only examine the origins of the characteristic peak structures of {\it case 1} and {\it case 2}, which are the candidates for the polaron state of K-TCNQ.

\section{Origins of Midgap States}

\subsection{\it Case 1}

\begin{figure}[hbt]
\begin{center}
\includegraphics[width=6.0cm,clip]{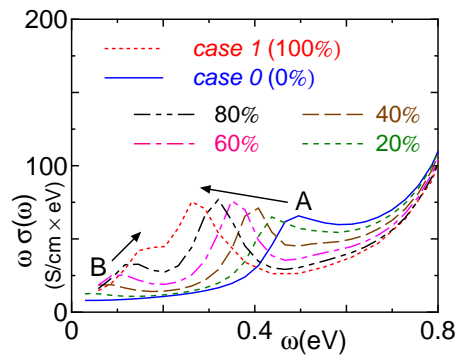}
\end{center}
\caption{(Color online) Spectral functions for several lattice configurations (see the text).}
\label{fig_case1}
\end{figure}

%First, let us examine the {\it case 1}.
In Fig.~\ref{fig_case1}, we plot $\omega\sigma(\omega)$ in {\it case 1}, and that of the unrelaxed lattice configuration ({\it case 0}), where $\bm{u}$ and $\bm{v}$ are the same as those at half filling.  Further we show $\omega\sigma(\omega)$ for several intermediate lattice configurations between {\it case 0} and {\it case 1}, defined as
\begin{equation}
 \bm{u}(x) = \bm{u}({\rm case \ 0}) + x [\bm{u}({\rm case \ 1})-\bm{u}({\rm case \ 0})],
\end{equation}
where the degree of relaxation is denoted by $x$, and in Fig.~\ref{fig_case1}, $x$ is shown in percentage.

The most intriguing is that a midgap peak, termed (A), appears at 0.5eV in {\it case 0}.  Since no analog is present in noninteracting band insulators, this midgap peak is characteristic to the 1D dimerized Mott insulators.  The peak (A) shifts toward lower energy as the lattice configuration approaches the completely $u$-relaxed one (100\%).  We also note that the other midgap peak, termed (B), appears in the lowest energy region for the slightly relaxed configuration ($x=20\%$).  The peak (B) moves towards higher energy as the relaxation proceeds, and finally becomes a shoulder at $\bm{u}(100\%)$.

\begin{figure}[hbt]
\begin{center}
\includegraphics[width=6.0cm,clip]{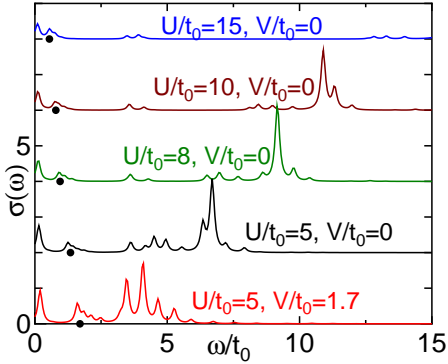}
\end{center}
\caption{(Color online) Optical conductivity spectra for several sets of ($U,V$) with unrelaxed lattice configuration ({\it case 0}). The results are obtained with Lanczos diagonalization for $N=12$ systems with 7 electrons and the OBC imposed. The broadening parameter is set to $\epsilon=0.1t_0$.  The filled circles show the singlet-triplet gap given by eq.~(\ref{eq_sgap})}
\label{fig_case0}
\end{figure}

To clarify the origins of these midgap peaks, we calculate the optical conductivity spectra of unrelaxed systems with varying $U$ and $V$, which are shown in Fig.~\ref{fig_case0}.  We can see three different classes of midgap peaks.  One is the metallic low-energy component in the lowest frequency region.  This is the Drude ``precursor'' in the OBC case~\cite{fye}.
Another excitation is observed around $\omega\sim 3t_0$, which is well separated from the CT band for $U/t_0\ge 8$.
This peak is caused by the intradimer CT excitation at the carrier site, while the carrier is mobile.~\cite{maeshima3}  In ref.~\citen{maeshima3}, we argued that this CT excitation is a possible origin of the midgap peak in K-TCNQ.  However, the present estimation of the model parameters for K-TCNQ ($U/t_0=5,V/t_0=1.7$) demonstrates that the corresponding peak is hidden by the CT excitations around $\omega\sim 4t_0$.

Now we focus on the other class around $\omega/t_0=1-2$.  For the parameter set corresponding to K-TCNQ, this peak is located at $\omega/t_0\sim 2$, which is identical to the midgap peak (A).
Thus this midgap peak (A) can appear generally in 1D dimerized Mott insulators.
It should be noted that its location is quite sensitive to $U$ and $V$; the excitation energy increases as $U$ decreases and as $V$ increases, implying that the midgap peak (A) is caused by spin excitations.

\begin{figure}[hbt]
\begin{center}
\includegraphics[width=6.0cm,clip]{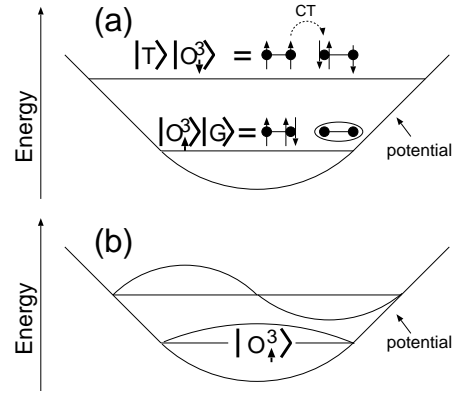}
\end{center}
\caption{Schematic energy diagram for relevant midgap states (a) and (b) causing peaks (A) and (B), respectively. }
\label{fig_midgap}
\end{figure}

In fact, it can be shown that the excitation causing the peak (A) includes a singlet-triplet excitation.
To show this, we use the analytical expression in the decoupled-dimer limit.~\cite{maeshima3}
In the 0-th order approximation, the ground state of the system is represented by the direct product of the singlet-dimer states;
\begin{equation}
|\psi_0^0\rangle = |G\rangle_{0} \otimes |G\rangle_{1} \otimes \cdots \otimes |G\rangle_{N_d-1},
\end{equation}
where $|G\rangle_l,(l=0,\cdots,N_d-1)$ is the singlet-dimer state on the $l$-th dimer, and $N_d=N/2$.
Then the one-electron-doped system is denoted by the linear combination of a direct product given by
\begin{equation}
|G\rangle_{0} \otimes |G\rangle_{1} \otimes \cdots \otimes |O^3_\uparrow\rangle_n    \otimes\cdots \otimes |G\rangle_{N_d-1}.
\label{eq_o3}
\end{equation}
Here, $|O^3_\uparrow\rangle_n$ has one more electron, and is defined as
\begin{equation}
|O^3_{\sigma}\rangle_n = \frac{1}{\sqrt{2}} ( c^\dagger_{2n\bar{\sigma}} c^\dagger_{2n+1\sigma}  c^\dagger_{2n\sigma} 
+ c^\dagger_{2n+1\bar{\sigma}}  c^\dagger_{2n+1\sigma}  c^\dagger_{2n\sigma}  ) |0\rangle,
\end{equation}
with spin $\sigma$. Then we can construct a CT excitation via a interdimer CT process to obtain
\begin{equation}
|G\rangle_{0} \otimes  \cdots \otimes |T\rangle_n\otimes |O^3_\downarrow\rangle_{n+1}   \otimes\cdots \otimes |G\rangle_{N_d-1},
\label{eq_t}
\end{equation}
where
\begin{equation}
 |T\rangle_n = c^\dagger_{2n\uparrow} c^\dagger_{2n+1\uparrow}|0\rangle.
\end{equation}
A schematic picture of this excitation is given in Fig.~\ref{fig_midgap} (a).  Comparing eq.~(\ref{eq_o3}) and eq.~(\ref{eq_t}), we find that the corresponding excitation energy $\Delta$ is equal to the singlet-triplet gap given by 
\begin{equation}
 \Delta = \frac{V-U}{2} + \sqrt{ ( U-V)^2/4 + 4t_0^2(1+\delta)^2}.
\label{eq_sgap}
\end{equation}
In Fig.~\ref{fig_case0}, we plot $\Delta$ for each parameter set and find that $\Delta$  explains the excitation energy of the midgap peak (A).

\begin{figure}[hbt]
\begin{center}
\includegraphics[width=6.0cm,clip]{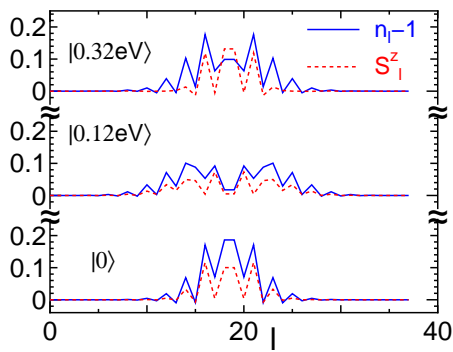}
\end{center}
\caption{(Color online) Charge and spin distributions in several states for $x=80\%$ with $N=38$.}
\label{fig_density}
\end{figure}

The existence of two states, $|T\rangle_n$ and $|O^3_\downarrow\rangle_m$, is confirmed in the following analysis.  If the state causing the peak (A) has these states, the doubly occupied site and the up spin must be separated ($n \ne m$).  This is completely different from the states that include only $|O^3_\uparrow\rangle_n$, where the doubly occupied site and the up spin are located on the same dimer $n$.  To demonstrate this spin-charge ``separation'', we calculate the corresponding correction vector~\cite{iwano2},
\begin{equation}
 |\omega\rangle= A \frac{1}{({\cal H} - E_0 -\omega)^2 + \epsilon^2} J|\psi_0\rangle,
\label{eq:psi_w}
\end{equation}
with normalization constant $A$.
Figure~\ref{fig_density} shows the charge and spin density profile of the correction vectors
for a nearly completely relaxed configuration (80\%).
Here, we used the broadening parameter $\epsilon=0.16t_0$, whose result is found to be almost identical to that with $\epsilon=0.08t_0$ (not shown).
%, where the lower midgap peak (B) at 0.12 eV and the higher peak (A) at 0.32 eV is sufficiently separated.
We can see that the spin density of the state $|0.32{\rm eV}\rangle$ concentrates on the center of the lattice deformation and the charge density has a double peak beside the center, which shows that the spin and charge densities are separated as discussed above. % and the triplet $|T\rangle$ exists on the center of the lattice deformation

By contrast, the spin and charge densities of the state $|0.12{\rm eV}\rangle$ have similar distributions; both have  double peaks and nodes at the center.  Since the lowest state $|0\rangle$ has no node, the state causing the midgap peak (B) is the second lowest state made of the elementally excitation $|O^3_\uparrow\rangle$, as shown in Fig.~\ref{fig_midgap} (b).

\subsection{\it Case 2}

\begin{figure}[hbt]
\begin{center}
\includegraphics[width=6.0cm,clip]{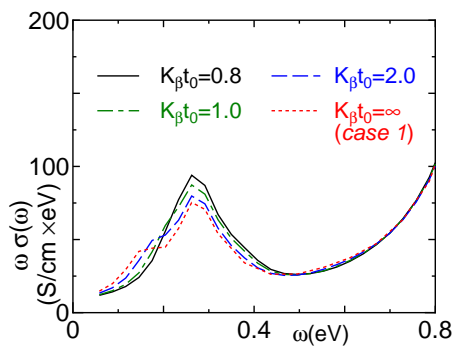}
\end{center}
\caption{(Color online) Spectral functions in {\it case 2}.}
\label{fig_case2}
\end{figure}

\begin{figure}[hbt]
\begin{center}
\includegraphics[width=6.0cm,clip]{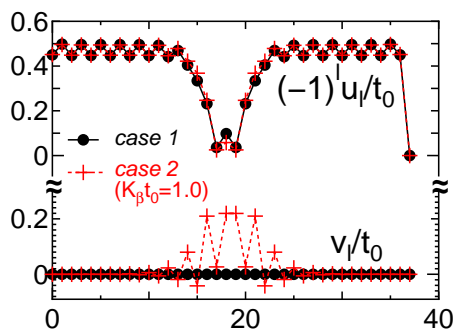}
\end{center}
\caption{(Color online) Lattice configurations for {\it case 1} and {\it case 2} with $K_\beta t_0=1.0$ for $N=38$ systems.}
\label{fig_lattice_case2}
\end{figure}

Now, we turn our attention to {\it case 2}.  Figure~\ref{fig_case2} shows the spectral function $\omega \sigma(\omega)$ in {\it case 2} for several values of $K_\beta$.  We can see that the shoulder structure (B) for $K_\beta t_0=\infty$ approaches the main peak (A) as $K_\beta$ decreases.  Finally, these seem to merge into a single peak.  This is because the excitation gap between the lowest polaron state and the second lowest state increases as the one-electron potential due to the intramolecular deformation $\bm{v}$ becomes deep, as shown in Fig.~\ref{fig_lattice_case2}.  By contrast, the intermolecular deformation $\bm{u}$ changes slightly, keeping the singlet-triplet gap $\Delta$ almost constant.

\section{Summary and Discussion}

We have studied polaronic midgap states in the 1D extended Hubbard model with Peierls and Holstein types of electron-phonon couplings to clarify the origin of the midgap peak in K-TCNQ.  Our main conclusion is that the relaxation of the intermolecular lattice distortion $\bm{u}$ is necessary to reproduce the observed midgap peak at 0.3 eV.

The midgap peak is numerically reproduced in the spectral function $\omega \sigma(\omega)$ for two polaronic lattice configurations.  One has relaxed $\bm{u}$ and unrelaxed intramolecular vibration mode $\bm{v}$ ({\it case 1}), and the other has relaxed $\bm{u}$ and relaxed $\bm{v}$ ({\it case 2}).
The former configuration is found to exhibit an additional shoulder structure on the lower energy side of the midgap peak.  In this case, we have clarified the origins of the shoulder and the midgap peak.

The shoulder structure is caused by the excitation from the lowest bound state to the second-lowest state, both of which consist of the elementary excitation $|O^3_\uparrow\rangle$.  This is similar to the polaron-based lower midgap peak of usual 1D band insulators, where the peak stems from the excitation from the lowest bound state to a higher state of introduced carriers (holes or electrons).~\cite{heeger,tagawa}

By contrast, the midgap peak is characteristic to 1D dimerized Mott insulators.  The peak corresponds to the interdimer CT excitation from the state $|O^3_\uparrow\rangle$ to a superposition of the state $|O^3_\downarrow\rangle$ and the triplet dimer state $|T\rangle$. Its excitation energy is shown to exhibit spin-excitation-like $(U,V)$-dependence.

Our results provide some suggestions to the photoinduced inverse spin-Peierls ``transition.''~\cite{KTCNQ1,KTCNQ2,KTCNQ3}  First, the polaronic lattice relaxation proceeds within the experimental time resolution ($=$150fs).
This implies that the phonon motion to generate the polaronic state is much faster than the observed coherent oscilation.~\cite{KTCNQ2}  To discuss the speed of the phonon motion, we here evaluate the bare phonon frequency.  The experimentally observed dimerization length ($\sim$ 0.165\AA~)\cite{konno} and the dimerization strength of the transfer integral ($\delta t_0=0.46\times 0.291$eV) lead to the e-ph coupling constant $\alpha\sim 0.81$ eV/\AA.  Then the unrenormalized elastic constant $K_\alpha$ is obtained to be 3.6 eV/\AA$^2$.
These values for physical parameters give the bare phonon frequency $\omega=2\sqrt{K_\alpha/M_{\rm TCNQ}}\sim 2.6\times 10^{13}{\rm s}^{-1}$, where $M_{\rm TCNQ}\sim3.4\times 10^{-25}{\rm kg}$ is the mass of a TCNQ molecule.
 Thus the speed of the bare phonon ($\sim$ sub ps) is found to be comparable to the experimental resolution time. It remains unresolved whether the bare phonon frequency determines the speed of the polaron formation or not.

%This is opposite to the our previous speculation.~\cite{maeshima3}  The speculation was deduced from the inconsistency between the fast formation of the midgap peak and the slow phonon motion.~\cite{KTCNQ2}

Second, the photo-induced melting of the dimerization is explained by the ``impurity (polaron) doping'' picture~\cite{KTCNQ2}, not by the mobile-carrier picture,~\cite{maeshima3} where the mobile photodoped carriers destroy the singlet dimerization of the dimerized phase.
Third, one carrier can weaken the lattice dimerization roughly over 10 molecules as shown in Fig.~\ref{fig_lattice_case2}. Assuming that interaction between doped carriers is negligible, we estimate that the dimerization order decreases linearly with respect to doping concentration at least up to 0.1 carrier per molecule.  This is equivalent to 0.05 photon per molecule, which is equal to the photon density where the $\Delta R/R$ at 0.71 eV saturates [see Fig. 2(d) of ref.~\citen{KTCNQ2}].

Our estimation of the model parameters for K-TCNQ has demonstrated that K-TCNQ does not show the absorption peak caused by the conversion of photogenerated elementary excitations, which was proposed as the origin of the midgap peak.~\cite{maeshima3}  This absorption for K-TCNQ is hidden by the CT excitation because the CT gap is nearly equal to the location of this absorption.

The existence of this conversion can be confirmed in other quasi-1D materials that have a larger CT gap.  For example, an organic radical crystal, 1,3,5-trithia-2,4,6-triazapentalenyl (TTTA) is a candidate of these materials.~\cite{TTTA2}  In ref.~\citen{maeshima4}, we have estimated model parameters for TTTA as $\delta=0.4$ and $U/t_0=10,V/t_0=1.7,t_0=0.235$ eV.  Since this material has relatively large $U/t_0$, the absorption caused by the conversion can be well separated from the CT band.

%We also note that the absorption can appear after photoirradiation as long as the quasi-1D lattice structure of TTTA crystal is maintained. ; and the excitation  The former state , suggesting that quasi-1D materials could generally exhibit a similar midgap peak. , such as TTTA,~cite{TTTA2}  before the quasi-1D lattice structure is destroyed by photoirradiation.~\cite{TTTA1,takeda}

%---------------------------------------------------------------------------------------------
\section*{Acknowledgments}

The authors are grateful to Prof. H.~Okamoto for enlightening discussions.  This work was supported by Grants-in-Aid for Creative Scientific Research (No.~15GS0216), for Scientific Research on Priority Area ``Molecular Conductors'' (No.~15073224), for Scientific Research (C) (No.~19540381), and Next Generation SuperComputing Project (Nanoscience Program), from the Ministry of Education, Culture, Sports, Science and Technology, Japan.  Some of numerical calculations were carried out on Altix3700 BX2 at YITP in Kyoto University, and on TX-7 and PRIMEQUEST at Research Center for Computational Science, Okazaki, Japan.

%-------------------------------------------------------------------------------------------------------------

\end{document}